\documentclass[aps,preprint]{revtex4}%
\usepackage{amsfonts}
\usepackage{amsmath}
\usepackage{amssymb}
\usepackage{graphicx}%
\begin{document}
\title[Blackbody Radiation and Relativity]{Connecting Blackbody Radiation, Relativity, and Discrete Charge in Classical Electrodynamics}
\author{Timothy H. Boyer}
\affiliation{Department of Physics, City College of the City University of New York, New
York, NY 10031}

\begin{abstract}
It is suggested that an understanding of blackbody radiation within classical
physics requires the presence of classical electromagnetic zero-point
radiation, the restriction to relativistic (Coulomb) scattering systems, and
the use of discrete charge. \ The contrasting scaling properties of
nonrelativistic classical mechanics and classical electrodynamics are noted,
and it is emphasized that the solutions of classical electrodynamics found in
nature involve constants which connect together the scales of length, time,
and energy. \ Indeed, there are analogies between the electrostatic forces for
groups of particles of discrete charge and the van der Waals forces in
equilibrium thermal radiation. \ The differing Lorentz- or
Galilean-transformation properties of the zero-point radiation spectrum and
the Rayleigh-Jeans spectrum are noted in conection with their scaling
properties. \ Also, the thermal effects of acceleration within classical
electromagnetism are related to the existence of thermal equilibrium within a
gravitational field. \ The unique scaling and phase-space properties of a
discrete charge in the Coulomb potential suggest the possibility of an
equilibrium between the zero-point radiation spectrum and matter which is
universal (independent of the particle mass), and an equilibrium between a
universal thermal radiation spectrum and matter where the matter phase space
depends only upon the ratio $mc^{2}/k_{B}T$. \ The observations and
qualitative suggestions made here run counter to the ideas of currently
accepted quantum physics.

\end{abstract}
\maketitle

\section{Introduction}

Although blackbody radiation, relativity, and discrete charge are rarely
connected in the physics literature, they are intimately connected in nature.
\ Thus, for example, Planck's blackbody spectrum appears as the equilibrium
spectrum associated with uniform proper acceleration through a
Lorentz-invariant spectrum of random classical electromagnetic
radiation.\cite{thermal}\cite{thermal2}\cite{thermal3}\cite{thermal4}
\ Furthermore, the charge of the electron and Stefan's blackbody radiation
constant can be combined to give a universal dimensionless constant. \ On the
other hand, the many attempts to understand blackbody radiation within
classical physics using nonrelativistic statistical mechanics or
nonrelativistic mechanical scattering systems with a small-charge coupling
limit all lead to the Rayleigh-Jeans spectrum with its ultraviolet divergence.
On this account most physicist believe that the blackbody radiation spectrum
arises from a charge of arbitrary size in any (nonrelativistic) mechanical
potential because of Boltzmann statistics and the quantum nature of energy
exchanges. \ Here we reexamine the classical electromagnetic description of
radiation equilibrium. \ We suggest that the observed Planck spectrum of
blackbody radiation may have nothing to do with energy quanta and everything
to do with the symmetries of relativistic classical electron theory with
discrete charges.\cite{conform}

The physicists who investigated blackbody radiation near the beginning of the
twentieth century were unfamiliar with the implications of special relativity
and so interpreted "classical physics" to mean "nonrelativistic classical
mechanics." \ Furthermore, the discrete electronic charge was viewed (and
still is today) as a curiosity unrelated to blackbody radiation. \ Thus the
normal modes of oscillation of the electromagnetic field were treated as
mechanical waves using nonrelativistic classical statistical
mechanics\cite{Eisberg} or the electric dipole oscillator in contact with the
radiation field was treated by nonrelativistic classical statistical
mechanics\cite{Lavenda} or the scatterer of radiation was a nonrelativistic
classical mechanical scatterer.\cite{nonlinear} Researchers around 1900
wondered how the variety of nonrelativistic mechanical systems could possibly
lead to the observed universal spectrum for radiation equilibrium which was
unrelated to the details of the matter \ producing the equilibrium. \ Finally
around 1910 it was noted that the Rayleigh-Jeans spectrum always seemed to
appear from treatments using nonrelativistic classical mechanics, and moreover
nonrelativistic classical mechanics does not include any fundamental constant
which could lead to a departure from the Rayleigh-Jeans spectrum.\cite{Kuhn}
\ Indeed, the principles of nonrelativistic classical mechanics (involving
independent scalings of length, time, and energy) simply can not support a
fundamental constant like Stefan's constant $a_{s}$ connecting the energy
density $u$ of thermal radiation and the absolute temperature $T$,
$u=a_{s}T^{4}.$\ 

Today \textit{relativistic} physics is regarded as fundamental, not
nonrelativistic mechanics. \ In particular, the relativistic Coulomb
interaction between discrete point charges allows a separation between
particle mass and the particle phase space distribution which is not possible
for any other potential. \ Therefore it fits qualitatively with the Planck
spectrum of electromagnetic radiation as nonrelativistic mechanics does not.
\ Indeed, it is one of the ironies of the history of physics that blackbody
\ radiation, special relativity, and discrete electronic charge all came to
prominence at the beginning of the 20th century yet these were not connected.
\ Thus at the same time that Lorentz invariance was recognized as a symmetry
of electromagnetic waves, the interaction of radiation and matter was treated
by nonrelativistic mechanics for particles of arbitrarily small charge. \ It
was only half-a-century later that blackbody radiation and special relativity
began to be related in connection with the Lorentz invariance of zero-point
radiation\cite{Marshall}\cite{Boyer1975} and the thermal behavior associated
with uniform acceleration through zero-point radiation.\cite{thermal}%
\cite{thermal2}\cite{thermal3}\cite{thermal4} \ However, even today the
textbooks of modern physics hark back to the years of disconnection of a
century ago, while classical electromagnetic theory is taught as though
relativistic particle motion was not important and as thought electric charge
had no smallest value. \ It is our unproven suggestion that within classical
physics, the crucial appearance of the Rayleigh-Jeans spectrum or the Planck
spectrum has nothing to do with classical versus quantum physics but rather is
a reflection of the differing correlations allowed by nonrelativistic or
relativistic classical scattering systems with continuous or discrete charge.

\subsection{Outline of the Discussion}

This article is an attempt to exploring all the suggestive evidence for a
classical explanation of blackbody radiation and to understand why the
currently-accepted arguments are misleading. \ We start by pointing out the
contrasting scaling properties of nonrelativistic classical mechanics and
classical electrodynamics. \ It is emphasized that the solutions of classical
electrodynamics found in nature involve constants which connect together the
scales of length, time, and energy. \ Indeed, there are analogies between the
forces found in the electrostatics of discrete point charges and those found
between materials in equilibrium random classical radiation. \ Second we
consider equilibrium classical radiation and note the differing transformation
properties of the zero-point spectrum and the Rayleigh-Jeans spectrum. \ Also,
we remark on the appearance of thermal effects of acceleration within
classical electromagnetism and relate them to the existence of thermal
equilibrium in a gravitational field. \ Third, we note the unique scaling
properties of the Coulomb potential and the unique separation between the
phase space distribution and the particle mass. Fourth we discuss the
interaction between radiation and matter. \ We note that the Coulomb potential
with a discrete charge allows an interaction with random radiation which
connects the phase space of the matter with the phase space of the radiation
variables in a universal connection. \ Finally, we discuss our current
understanding of the blackbody radiation spectrum within classical physics.

\section{Scaling and Universal Constants}

The contradictions between nonrelativistic mechanics and electromagnetism can
be seen immediately from the contrasting scaling symmetries of nonrelativistic
classical mechanics as compared to classical electrodynamics with relativistic
particles of discrete charge. \ Thus attempts to explain blackbody radiation
based upon nonrelativistic classical statistical mechanics or nonrelativistic
classical scattering systems are doomed to failure. \ 

\subsection{Scaling for Nonrelativistic Mechanics}

Within nonrelativistic mechanics, length, time, and energy all scale
independently. \ The symmetries of classical mechanics allow separate
dilatation factors $\sigma_{l},$ $\sigma_{t},$ $\sigma_{E}$ for length, time
and energy, $l\rightarrow l^{\prime}=\sigma_{l}l$, $t^{\prime}\rightarrow
\sigma_{t}t,$ $E\rightarrow E^{\prime}=\sigma_{E}E,$ where the three separate
dilatation factors range over all positive real numbers. \ Thus for any
nonrelativistic mechanical system, there exists, in principle, a second system
which is twice as large, has a period three times as long, and contains four
times the energy. \ Since nonrelativistic mechanics allows independent
scalings of length, time, and energy, nonrelativistic mechanics can have no
fundamental constants connecting length, time, and energy. \ The existence in
the nineteenth century of independent standards of length, time, and energy
reflects the independent scalings found in nonrelativistic classical mechanics.

\subsection{Scaling for Classical Electromagnetism}

In the electromagnetic evidence accumulated during the last half of the
nineteenth century, this absence of any universality within nonrelativistic
classical mechanics stood in startling contrast with the appearance of a
universal wave speed $c=3\times10^{10}cm/$\textit{sec} in Maxwell's equations,
a universal energy-length-related constant $a_{s}/k_{B}^{4}=6.25\times
10^{64}(erg-cm)^{-3}$ (Stefan's constant divided by the fourth power of
Boltzmann's constant) for blackbody radiation, and a second
energy-length-related constant $e^{2}=2.304\times10^{-19}erg-cm$ corresponding
to a smallest electric charge $e$. \ Thus in contradiction to the separate
scalings found in nonrelativistic mechanics, classical electromagnetism has
the scales of length, time, and energy all connected. \ Maxwell's equations
themselves contain the speed of light $c$ in vacuum, and this fundamental
constant couples the scales of length and time. \ Thus if we find an
electromagnetic wave in vacuum with wavelength $\lambda$, then we know
immediately that the frequency $\nu$ of the wave is given by $\nu=c/\lambda$.
\ Furthermore, the solutions of Maxwell's equations found in nature during the
nineteenth century involve two other fundamental constants, one for radiation
and one for matter: Stefan's constant $a_{s}$ for blackbody radiation and a
smallest electric charge $e$. \ Stefan's constant $a_{s}$ divided by the
fourth power of Boltzmann's constant $k_{B}$ (this last connects absolute
temperature $T$ to energy) can be regarded as coupling together length and
energy through the relation for the electromagnetic thermal energy $U$ in a
cubic volume of side $l$ at temperature $T$ given by $U/(k_{B}T)^{4}%
=(a_{s}/k_{B}^{4})l$

$\ \ \ \ \ \ \ \ \ \ \ \ ^{3}$. \ Similarly, the smallest (nonzero) electric
charge $e$ couples the scales of length and energy for matter. \ Thus if two
smallest charges $e$ are separated by a distance $r$, then the electrostatic
potential energy is given by $U=e^{2}/r$. \ \ Within classical
electromagnetism, there is only one independent scaling in nature; the
dilatation symmetry allows only one scale factor $\sigma_{ltE^{-1}}$ giving
\begin{equation}
l\rightarrow l^{\prime}=\sigma_{ltE^{-1}}l,\text{ \ \ \ }t^{\prime}%
\rightarrow\sigma_{ltE^{-1}}t,\text{ \ \ \ }E\rightarrow E^{\prime}%
=E/\sigma_{ltE^{-1}}%
\end{equation}
where the dilatation factor $\sigma_{ltE^{-1}}$can assume all positive real
values. \ Given any classical electromagnetic system, there exists, in
principle, a second electromagnetic system where all the lengths, times, and
inverse energies are $\sigma_{ltE^{-1}}$ times as large.

\subsection{Electrostatic Energy-Length Scaling Due to Discrete Charge e}

The existence of a fundamental energy-length connection in electromagnetism
limits the possible energy of an electrostatic configuration. \ Thus if we are
told the number of elementary charges involved and the shape of the charge
distribution, then we know the product of the electrostatic energy times a
characteristic length of the distribution. \ Two elementary examples
immediately come to mind. \ A parallel-plate capacitor with square plates of
side $L$ separated by a distance $L/100$ and charged with $N$ elementary
charges (spread uniformly on each plate but of opposite sign for the two
plates) has an electrostatic energy $U$ where $UL=\{[1/(8\pi)][4\pi
Ne/L^{2}]^{2}L^{3}/100\}L=N^{2}e^{2}/50;$ a spherical conducting shell of
outer radius $a$ which contains $N$ elementary electric charges (spread
uniformly over the spherical shell) has an electrostatic energy $U$ where
$Ua=N^{2}e^{2}/2.$ \ In every case, the energy times the characteristic length
equals a shape-dependent geometrical factor times $N^{2}e^{2}.$ \ In every
case, there is a smallest nonvanishing energy when $N=1.$

\subsection{Electromagnetic Energy-Length Scaling Due to Stefan's Constant}

Now it is found in nature\cite{Sparnaay} that there are van der Waals forces
between macroscopic materials, and, at the absolute zero of temperature $T=0$,
these forces assume their smallest values for a given geometrical
configuration. \ It turns out that the smallest force at $T=0$ gives a
fundamental connection between energy and length which involves Stefan's
constant and is completely analogous to that noted above in electrostatics
where a minimum charge is involved.\cite{Casimirmodel} \ Thus if we are told
the shape of a distribution of conductors, then we know the product of the
electromagnetic energy times a characteristic length of the distribution at
absolute zero. \ Again two elementary examples come to mind \ At temperature
$T=0$, an uncharged parallel-plate capacitor\cite{Casimir} with square plates
of side $L$ separated by a distance $L/100$ has an energy $U$ where
$UL=-(\pi^{2}\times10^{6}/360)[120a_{s}/(\pi^{2}k_{B}^{4})]^{-1/3};$ a
spherical conducting shell\cite{sphere} of outer radius $a$ has an energy $U$
where $Ua=0.09[120a_{s}/(\pi^{2}k_{B}^{4})]^{-1/3}.$ \ In every case, the
energy times the characteristic length equals a shape-dependent geometrical
factor times $(a_{s}/k_{B}^{4})^{-1/3}.$

\subsection{Existence of a Dimensionless Universal Constant}

Since\ nature gives us two universal electromagnetic constants connecting the
scales of energy and length, it follows that their ratio must be a
dimensionless constant. \ Thus for example, the charge of the electron is
$e=4.80\times10^{-10}$ esu, so $e^{2}=2.304\times10^{-19}erg-cm$ while
Stefan's constant is $a_{s}=7.56\times10^{-15}erg\cdot cm^{-3}\cdot K^{-4}$
and Boltzmann's constant is $k_{B}=1.38\times10^{-16}erg\cdot K$. Accordingly,
we note that%

\begin{equation}
e^{2}(a_{s}/k_{B}^{4})^{1/3}=0.00063487
\end{equation}
\ This ratio involving nineteenth-century physical constants is not usually
presented in the physics literature. \ It suggests the possibility that a
discrete electronic charge is connected to blackbody radiation.\cite{hbarc}

In the analysis to follow, we will frequently use the blackbody constant
$(a_{s}/k_{B}T)^{-1/3}.$ \ Therefore it is convenient to introduce Stefan's
second constant $b_{s}$ where\cite{bs}
\begin{equation}
b_{s}=\left(  \frac{\pi^{2}}{120}\frac{k_{B}^{4}}{a_{s}}\right)  ^{1/3}%
\end{equation}
The constant $b_{s}$ has the units of $energy\times length$, just the same as
$e^{2}.$

\section{Equilibrium Classical Radiation}

\subsection{Classical Zero-Point Radiation}

Pure classical electromagnetic radiation is a homogeneous solution of
Maxwell's equations. \ Classical radiation therefore contains the length-time
connection given by the wave speed $c$ in vacuum of Maxwell's equations, but
makes no connection between energy and length or between energy and time. For
example, an electromagnetic plane wave solution of Maxwell's equations in
vacuum of wavelength $\lambda$ must have a frequency $\nu$ given by
$\nu=\lambda/c$, but it may have any energy per unit volume associated with
the electric field amplitude $E_{0}.$ \ It follows that any fundamental
connection between energy and length involving electromagnetic radiation
within classical physics must come not from Maxwell's equations themselves but
from a fundamental boundary condition on Maxwell's equations. \ 

Nature indeed provides a fundamental boundary condition on Maxwell's
equations. \ All of the experimentally observed van der Waals
forces\cite{Sparnaay} between macroscopic objects can be described in terms of
classical electromagnetic forces due to random classical electromagnetic
radiation.\cite{temperature} \ The experimentally observed van der Waals
forces require that at temperature $T=0,$ there is present in space a
Lorentz-invariant spectrum\cite{Marshall} of random classical electromagnetic
radiation with an average energy $U_{\lambdabar}$ per normal mode of
wavelength $\lambda=2\pi\lambdabar$ given by
\begin{equation}
U_{\lambdabar}(0)=[(120/\pi^{2})(a_{s}/k_{B}^{4})]^{-1/3}/\lambdabar
=b_{s}/\lambdabar
\end{equation}
where $b_{s}$ is Stefan's second constant given in Eq.\ (3). \ The randomness
of the radiation can be described in terms of random phases for the
radiation\cite{Rice} and the radiation itself can be described in terms of
normal modes with action-angle variables and an associated probability
function $P_{\lambdabar}$ for the action variable $J_{\lambdabar}$ of the
radiation mode of wavelength $\lambdabar$\cite{Power}\cite{adiabatic}%
\begin{equation}
P_{\lambdabar}(J_{\lambdabar},0)=\frac{c}{b_{s}}\exp\left[  -\frac
{J_{\lambdabar}c}{b_{s}}\right]
\end{equation}
We notice in Eq. (5) that the probability function $P_{\lambdabar
}(J_{\lambdabar},0)$ at temperature $T=0$ is the same for every radiation mode
independent of the wavelength $\lambdabar$ of the mode.

We expect that the entropy associated with any classical system should be
related to the probability distributions of its action variables since the
entropy is related to the probability distribution on phase space. \ The
action-angle variables of multiply periodic systems give a natural division of
phase space. \ The distribution in the angle variables is uniform, so that the
only probability distribution of relevance is the action variables. \ By the
third law of thermodynamics, we expect the thermodynamic entropy to vanish at
$T=0.$ \ Therefore the probability distribution at $T=0$ given in Eq. (5)
corresponds to zero entropy for each radiation \ mode.\cite{entropy}

\subsection{Thermal Radiation}

At finite temperature, the observed spectrum of blackbody radiation (including
the zero-point radiation required by the observed van der Waals forces) can be
written as an energy $U_{\lambdabar}(T)$ per normal mode%
\begin{equation}
U_{\lambdabar}(T)=\frac{b_{s}}{\lambdabar}\coth\left(  \frac{b_{s}}{\lambdabar
k_{B}T}\right)
\end{equation}
with the probability function $P_{\lambdabar}$ for the action variable
$J_{\lambdabar}$ of \ a radiation normal mode of wavelength $\lambdabar$
becoming a function of temperature,\cite{adiabatic}%
\begin{equation}
P_{\lambdabar}(J_{\lambdabar},\lambdabar T/b_{s})=\frac{c}{b_{s}\coth
[b_{s}/(\lambdabar k_{B}T)]}\exp\left[  -\frac{J_{\lambdabar}c}{b_{s}%
\coth[b_{s}/(\lambdabar k_{B}T)]}\right]
\end{equation}

Even at finite temperature, the van der Waals forces for a conducting-walled
container due to random classical radiation still hold a strong analogy with
the electrostatic forces due to discrete charges in their dependence upon
dimensionless and scaling parameters. \ \ We saw above that an electrostatics
problem was uniquely specified by giving the number of elementary charges, the
shape of the charge container holding uniformly spaced charges, and one
scale-determining length. \ In the thermal radiation problem for van der Waals
forces, the pure number $N$ (number of elementary charges) of the
electrostatics problem is replaced by the pure number $S/k_{B}$ corresponding
to the entropy of the radiation in the container. \ (As usual, we have removed
the inessential unit of temperature by dividing out Boltzmann's
constant.)\ \ Thus nature shows that thermal radiation in a container of given
shape (specified by dimensionless parameters) is determined by exactly two
parameters; viz, the volume $V$ (which corresponds to setting the length scale
$l=V^{1/3}$ of the container of given shape) and the scale-independent entropy
$S/k_{B}.$ \ All the other parameters are now fixed. \ For thermal radiation
in a large spherical container, the radiation temperature is $T=[3S/(4a_{s}%
V)]^{1/3}$, the total thermal radiation energy in the container is
$U=a_{s}V[3S/(4a_{s}V)]^{4/3}$ $=a_{s}T^{4},$ the energy per normal mode
$U_{\lambdabar}$ in the long wavelength (low frequency) modes is given by
$U_{\lambdabar}=k_{B}[3S/(4a_{s}V)]^{1/3}=k_{B}T,$ and the wavelength where
the thermal radiation spectrum has its maximum is given by Wien's displacement
law $\lambda_{\max}=const\times\lbrack4a_{s}V/(3S)]^{1/3}=const/T.$ \ 

Now classical electromagnetism is invariant under $\sigma_{ltE^{-1}}$ scaling
symmetry and even under conformal symmetry.\cite{gamma} \ Therefore an
adiabatic change in the volume of the container (while maintaining its shape)
is the same as a $\sigma_{ltE^{-1}}$ change of scale. \ For the electrostatic
situation, the energy $U$ changes as the scaling length $l$ changes while the
number of elementary charges remains fixed. \ For the radiation case, the
thermal energy in the container changes as the radius of the container changes
while the scale-invariant entropy $S/k_{B}$ is unchanged. If we imagine a
spherical container, the change in the radius of the container $r\rightarrow
r^{\prime}=\sigma_{ltE^{-1}}r$, leads to a consistent change of volume
$V\rightarrow V^{\prime}=\allowbreak\sigma_{ltE^{-1}}^{3}V,$ temperature
$T\rightarrow T^{\prime}=T/\sigma_{ltE^{-1}}$, and energy $U\rightarrow
U^{\prime}=U/\sigma_{ltE^{-1}},$ while the entropy is unchanged $S\rightarrow
S^{\prime}=S.$ \ All the laws of blackbody radiation hold both before and
after the scale change. \ The probability function $P_{\lambdabar
}(J_{\lambdabar},T)\rightarrow P_{\lambdabar^{\prime}}(J_{\lambdabar^{\prime}%
},T^{\prime})$ in Eq. (7) is unchanged since the product $\lambdabar
k_{B}T=(\sigma_{ltE^{-1}}\lambdabar)(k_{B}T/\sigma_{ltE^{-1}})=\lambdabar
^{\prime}k_{B}T^{\prime}$ is unchanged under the scale change or adiabatic
change, and this invariance is appropriate for the invariance of the radiation entropy.

The equivalence between adiabatic change in the size of a container (which
retains its shape) of electromagnetic energy and a $\sigma_{ltE^{-1}\text{ }}%
$scale transformation of the situation is something that can hold only for
pure electromagnetic fields and not when particle masses $m$ are involved.
\ We do not think of particles in a container as changing mass $m$ during an
adiabatic change, whereas the the numerical value of mass is rescaled under a
$\sigma_{ltE^{-1}}$ scale change, $m\rightarrow m^{\prime}=m/\sigma_{ltE^{-1}%
}$.

\subsection{Electric Field Correlation Functions in the High- and
Low-Temperature Limits}

In the high-frequency or short-wavelength limit, the blackbody radiation
spectrum (6) involves $\coth[x]\rightarrow1$ for large $x$ and goes over to
the zero-point energy limit%
\begin{equation}
U_{\lambdabar}(0)=U_{\omega}(0)=b_{s}/\lambdabar=(b_{s}/c)\omega
\end{equation}
In the long-wavelength or low-frequency limit, the blackbody spectrum (6)
involves $\coth[x]\rightarrow1/x$ for small $x$ and goes over to the the
Rayleigh-Jeans equipartition energy%

\begin{equation}
U_{RJ\lambdabar}=U_{RJ\omega}=k_{B}T
\end{equation}
This latter spectrum corresponds to the energy (with associated entropy) of
traditional nonrelativistic classical statistical mechanics per normal mode.
\ It is interesting to see the electromagnetic field correlation functions for
these two limiting spectra\ (8) and (9), and to note the role played by the
speed of light $c.$ \ The electric field correlation function for the
zero-point spectrum of random radiation is given by\cite{conformal}%
\begin{equation}
<E_{i}(\mathbf{r},t)E_{j}(\mathbf{r}^{\prime},t^{\prime})>_{T=0}=\left(
\frac{\delta_{ij}}{c^{2}}\frac{\partial}{\partial t}\frac{\partial}{\partial
t^{\prime}}-\frac{\partial}{\partial x_{i}}\frac{\partial}{\partial
x_{j}^{\prime}}\right)  \left(  \frac{-2b_{s}/\pi}{c^{2}(t-t^{\prime}%
)^{2}-(\mathbf{r-r}^{\prime})^{2}}\right)  \text{ }%
\end{equation}
The other electromagnetic field correlation functions in the four-tensor
expression $\left\langle F^{\mu\nu}(x)F^{\mu^{\prime}\nu^{\prime}}(x^{\prime
})\right\rangle $ can be obtained by changing the space and time indices in
Eq. (10).\cite{conformal} \ The electric field correlation function involves
space and time derivatives of the Lorentz-invariant spacetime interval
$c^{2}(t-t^{\prime})^{2}-(\mathbf{r-r}^{\prime})^{2}$ between the field points
$(\mathbf{r,}t)$ and $(\mathbf{r}^{\prime},t^{\prime}).$ This indeed denotes
the Lorentz-invariant character of the random radiation spectrum. \ Now we
expect, but have not proved, that relativistic scattering systems which
themselves maintain the Lorentz invariance of spacetime intervals will leave
this spectrum invariant. \ We suggest that it is unreasonable to expect that
\textit{nonrelativistic} mechanical scattering systems, which do not share the
Lorentz symmetry of the zero-point spectrum, should leave the
Lorentz-invariant zero-point spectrum unchanged.

The electric field correlation function for the Rayleigh-Jeans spectrum takes
the form\cite{RJ}%

\begin{equation}
<E_{i}(\mathbf{r},t)E_{j}(\mathbf{r}^{\prime},t^{\prime})>_{RJ}=\left(
\frac{\delta_{ij}}{c^{2}}\frac{\partial}{\partial t}\frac{\partial}{\partial
t^{\prime}}-\frac{\partial}{\partial x_{i}}\frac{\partial}{\partial
x_{j}^{\prime}}\right)  \frac{k_{B}T}{\left\vert \mathbf{r-r}^{\prime
}\right\vert }\theta(\left\vert \mathbf{r-r}^{\prime}\right\vert -c\left\vert
t-t^{\prime}\right\vert )\text{ }%
\end{equation}
We notice immediately that this correlation function depends not only on the
Lorentz-invariant spacetime interval $c\left\vert t-t^{\prime}\right\vert
-\left\vert \mathbf{r-r}^{\prime}\right\vert $ but also on the
Galilean-invariant interval $\left\vert \mathbf{r-r}^{\prime}\right\vert .$
\ Thus the right-hand side is Galilean invariant but not Lorentz invariant.
\ We expect that the spectrum will be preserved by electromagnetic scattering
systems which maintain the invariance of the nonrelativistic invariant spatial
interval. \ This is seen in several scattering calculations.\cite{nonlinear}
\ The correlation function in Eq. (11) vanishes for time-like separations, an
aspect which appears from the Lorentz-covariance of the radiation itself, not
from the information carried in the radiation spectrum . \ The only
correlations in time involve $\delta$-function correlations and so involve no
connection between energy and time, as is true in nonrelativistic mechanics.

\subsection{Limiting Field Correlations Functions as $c\rightarrow\infty$}

In order to emphasize the distinction between zero-point radiation and the
Rayleigh-Jeans spectrum, we will consider the limit $c\rightarrow\infty$ so as
to eliminate $c$ from Eqs. (10) and (11). \ The $c\rightarrow\infty$ limit is
possible only when $t=t^{\prime}.$ \ Then the zero-point radiation correlation
function (10) becomes%
\begin{equation}
<E_{i}(\mathbf{r},t)E_{j}(\mathbf{r}^{\prime},t)>_{T=0}=\left(  \frac
{\partial}{\partial x_{i}}\frac{\partial}{\partial x_{j}^{\prime}}\right)
\left(  \frac{-2b_{s}/\pi}{(\mathbf{r-r}^{\prime})^{2}}\right)
\end{equation}
while that for the Rayleigh-Jeans spectrum becomes%
\begin{equation}
<E_{i}(\mathbf{r},t)E_{j}(\mathbf{r}^{\prime},t)>_{RJ}=\left(  \frac{\partial
}{\partial x_{i}}\frac{\partial}{\partial x_{j}^{\prime}}\right)  \frac
{-k_{B}T}{\left\vert \mathbf{r-r}^{\prime}\right\vert }%
\end{equation}
The different functional dependence upon the the spatial interval $\left\vert
\mathbf{r-r}^{\prime}\right\vert $ seen in Eqs. (12) and (13) is a reminder of
the coupling between energy and length in zero-point radiation and the
complete decoupling between energy and length found in the Rayleigh-Jeans
spectrum. \ Thus the electric field (think of $E=e/r^{2}$) scales as two
inverse powers of the dilatation factor $\sigma_{ltE^{-1}}.$ \ The left-hand
sides of Eqs. (12) and (13) scale as ($1/\sigma_{ltE^{-1}})^{4}$ and have
units of energy per unit volume. \ For the zero-point radiation spectrum in
Eq. (12), there are four powers of length on the right-hand side so that
$b_{s}$ must have units of $energy\times length$ and can be a fundamental
constant relating energy and length which is invariant under scaling by
$\sigma_{ltE^{-1}}.$ \ On the other hand, the Rayleigh-Jeans spectrum in Eq.
(13) has only three powers of length on the right-hand side, and therefore the
parameter $k_{B}T$ must have the units of $energy$ and is subject to scaling
by $\sigma_{ltE^{-1}},$ $k_{B}T\rightarrow k_{B}T^{\prime}=k_{B}%
T/\sigma_{ltE^{-1}}.$ \ Thus the energy parameter $k_{B}T$ can take on any
non-negative value. \ The Rayleigh-Jeans spectrum reflects the completely
independent scaling of energy and length which is typical of nonrelativistic
mechanics. \ Indeed, when van der Waals forces between macroscopic objects are
calculated using the Rayleigh-Jeans spectrum, one finds that there is no net
change in the electromagnetic energy with position of the
objects;\cite{conjectured} all the forces are associated with changes of
entropy through the Helmholtz free energy. \ This fits exactly with the
absence of any energy-length connection within the Rayleigh-Jeans spectrum.

\subsection{Thermal Effects of Acceleration within Classical Theory}

One of the surprising realizations of the last quarter of the twentieth
century was that the field correlation functions of Planck's blackbody
spectrum appear when a system undergoes uniform acceleration through
zero-point radiation.\cite{thermal}\cite{thermal2}\cite{thermal3}%
\cite{thermal4} \ Although this appearance without any apparent application of
statistical mechanics has provided a profound quandary for quantum physics, it
seems a natural result within classical electrodynamics with classical
zero-point radiation.\cite{quandry} \ 

The equivalence principle connects accelerations to gravitational phenomena,
while acceleration through classical zero-point radiation is found to connect
zero-point radiation with the Planck spectrum of thermal radiation. \ Thus it
seems relevant to consider our sketch of relativistic classical
electrodynamics with discrete charge in connection with gravitational effects.
\ The simplest system to consider is the Rindler frame\cite{Rindler} involving
a time-independent coordinate system where each spatial point undergoes a
uniform proper acceleration through Minkowski spacetime. \ \ We expect that
thermal radiation equilibrium can exist within a gravitational field. \ Thus
we expect thermal equilibrium to exist for the Rindler frame. \ Now the
Rindler frame involves uniform proper acceleration
\begin{equation}
\mathbf{a}=\widehat{k}a=\widehat{k}c^{2}/z
\end{equation}
for a spatial point a distance $z$ from the event horizon at $z=0.$ \ The
smallest density of random classical radiation is that given by zero-point
radiation. \ The essential aspect is its Lorentz-invariant spectrum which can
be written as%
\begin{equation}
U_{\omega}=const\times\omega
\end{equation}
for the energy $U_{\omega}$ of a normal mode of frequency $\omega,$ where
there is an arbitrary multiplicative constant $const.$ \ It has been
shown\cite{thermal2}\cite{thermal3}\cite{thermal4} that the correlation
function for the random classical electromagnetic fields as observed at a
fixed spatial point in the Rindler frame no longer corresponds to a
Lorentz-invariant spectrum but rather to a spectrum%

\begin{equation}
U_{\omega}=const\times\omega\coth(\pi\omega c/a)
\end{equation}
This result can be obtained by considering the uniform \ proper acceleration
of a harmonic electric dipole system of fixed angular frequency $\omega_{0}$
taken in the point dipole limit as the mass of the oscillator particle goes to
infinity. \ We notice that the expression for the radiation energy $U_{\omega
}$ at frequency $\omega$ involves the hyperbolic cosine function with a
dependence upon $\omega c/a$. \ The proper acceleration $a$ takes the place of
the temperature in the blackbody radiation spectrum.

For frequencies $\omega$ small compared to $a/c$, $\omega<<a/c,$ the spectrum
is proportional to the acceleration $a$ and independent of frequency, just
like the Rayleigh-Jeans spectrum in Eq. (9). \ For large frequencies
$\omega>>a/c,$ the spectrum is still the Lorentz-invariant spectrum (15) which
increases linearly with frequency $\omega.$ The result in Eq. (16) is entirely
classical, depends crucially on the Lorentz-invariance of the original
spectrum (15) in Minkowski spacetime, and has nothing to do with any
fundamental constant except the speed of light in vacuum $c$. \ The
multiplicative scale of the of the Lorentz-invariant spectrum is given by the
arbitrary constant here labeled "$const.$" \ If we choose the constant so as
to fit with the experimentally observed spectrum of classical zero-point
radiation, then $const=b_{s}/c$ and the associated temperature is given by
$k_{B}T=b_{s}a/(\pi c^{2}).$ \ The fluctuations of the radiation can be
obtained from the random phases of the radiation modes.\cite{Rice} \ 

From Wien's displacement theorem (which holds even in gravitational fields),
we know that thermal equilibrium at any temperature $T$ is of the form
$U_{\omega}=\omega f(\omega/T)$ or$\ U_{\lambdabar}=cf(c/\lambdabar
T)/\lambdabar$ where $f$ is a universal function. \ In an inertial frame, we
may take the limit $T\rightarrow0,$ and recover the zero-point spectrum given
in Eq. (4) or (15). \ In the Rindler frame involving uniform acceleration, the
function $f$ must follow from Eq. (16) which was found from the acceleration
through zero-point radiation. \ Thus comparing Eq. (16) with the
zero-temperature limit where the $const=b_{s}/c,$ we find that proper
acceleration $a$ through zero-point radiation corresponds to a lowest possible
temperature $T_{\min}$ in a gravitation field $a$ given by%
\begin{equation}
k_{B}T_{\min}=b_{s}a/(\pi c^{2})
\end{equation}
and any additional random radiation will increase the energy per normal mode
$U_{\omega}$ according to the functional behavior%
\begin{equation}
U_{\omega}=\frac{b_{s}}{c}\omega\coth\left(  \frac{(b_{s}/c)\omega}{k_{B}%
T}\right)
\end{equation}

\section{Behavior of Matter}

Classical electromagnetic radiation within a container with perfectly
reflecting walls will never come to thermal equilibrium. \ Rather, there must
be some interaction between radiation and matter which brings the radiation to
equilibrium.\cite{harmonic} \ The state of thermal equilibrium for the
radiation will reflect some fundamental aspects of the matter which causes the
equilibrium. \ This conviction was expressed clearly by Lorentz back in the
early 1900s when he writes,\cite{ponderable} "...we may hope to find in what
manner the value of this constant [$\lambda T=const$] is determined by some
numerical quantity that is the same for all ponderable bodies." \ Here we are
proposing that the discrete electric charge $e$ and use of relativistic
interactions are the crucial elements for matter. \ Relativistic
electromagnetic interactions begin with the Coulomb potential which involves
unique properties related to scaling,\cite{scal} phase space, and radiation
emission, and these unique properties encourage the possibility of a classical
explanation for blackbody radiation.

\subsection{$\sigma_{ltE^{-1}}$ Scaling for a Central Potential\ \ }

Suppose we consider a charged particle $e$ in a central potential of the form
$U(r)=-k/r^{n}$. \ Now under the $\sigma_{ltE^{-1}}$ scaling of classical
electromagnetism, the potential energy $U$ transforms as $U\rightarrow
U^{\prime}=U/\sigma_{ltE^{-1}}$ while the distance $r$ transforms as
$r\rightarrow r^{\prime}=\sigma_{ltE^{-1}}r$ so that $k^{\prime}/r^{\prime
n}=k^{\prime}/(\sigma_{ltE^{-1}}r)^{n}=(k/r^{n})/\sigma_{ltE^{-1}}$ and hence
we must have $k^{\prime}=\sigma_{ltE^{-1}}^{n-1}k.$ \ This means that if in
the collection of allowed systems there is a potential $U=-k/r^{n}$ with
strength $k,$ then there must also be a potential $U=-k^{\prime}/r^{n}$ of
strength $k^{\prime}=\sigma_{ltE^{-1}}^{n-1}k.$ \ Thus only in the case of the
Coulomb potential where $n=1$ (and $k=e^{2}$ has the units of $energy\times
length)$ do we have the possibility of a scale-independent coupling
$k^{\prime}=k.$ For all other potentials $U(r)=-k/r^{n},$ $n\neq1,$ the
coupling constant must allow all real values $k.$ \ Thus only for the Coulomb
potential do we preserve the $\sigma_{ltE^{-1}}$ scaling of classical
electromagnetic theory rather than being forced into the traditional scaling
of nonrelativistic mechanics with separate, independent scalings for energy
and length. \ Whereas a point charge of smallest charge $e$ at a distance $r$
in a Coulomb potential of elementary strength $e$ always has a potential
energy $U=e^{2}/r$ where $e$ is a universal value, we have no such information
for a particle at radius $r$ in a general potential $U(r)=-k/r^{n}$ because
the potential strength $k$ is freely changeable so that $U=k/r^{n}$ can be any
real number. \ Thus a fundamental connection between energy and length is
given by the elementary charge $e^{2}$ and a fundamental connection between
energy and time is given by $e^{2}/c.$ \ Accordingly within relativistic
theory, a particle of mass $m$ and smallest charge $e$ is connected to the
characteristic energy $mc^{2},$ the characteristic length $e^{2}/(mc^{2})\,,$
and the characteristic time $e^{2}/(mc^{3}).$

\subsection{Dependence of Orbit Speed on Action Variables \ }

In addition to allowing a unique, $\sigma_{ltE^{-1}}$ scale-independent
smallest coupling constant, the Coulomb potential also involves a unique
separation between orbital speed and mass for fixed angular momentum $J,$
which separation is not possible for any other central potential
$U(r)=-k/r^{n}.$ \ If we consider a point charge $e$ in a circular orbit of
radius $r$ in a central potential $U(r)=-k/r^{n},$ \ then the equation of
motion $\mathbf{F=}d\mathbf{p}/dt$ and the angular momentum $J$ are given by
\begin{equation}
m\gamma v^{2}/r=nk/r^{n+1}\text{ \ and\ \ \ \ }J=mr\gamma v
\end{equation}
while the system energy is given by $H=m\gamma c^{2}-k/r^{n}.$ \ If we connect
the equation of motion and the angular momentum $J$ in Eq. (19) so as to
eliminate the orbital radius $r$, then we have $m\gamma v^{2}=nk(m\gamma
v/J)^{n}$ or $\gamma^{1-n}v^{2-n}=nkm^{n-1}/J^{n}.$ \ Thus only for $n=1$
(corresponding to the Coulomb potential $k=e^{2})$ does the particle mass $m$
disappear from this last equation so that we have the orbital speed $v$ of the
particle determined solely by the action variable $J$, $v=e^{2}/J.$ \ For
every other potential $U(r)=-k/r^{n}$, the orbital speed $v$ is determined by
\textit{both} $J$ and the product $km^{n-1}$ where $k$ (as seen above) can not
be a universal constant but must be a free scaling parameter. \ 

\subsection{Relativistic Coulomb Motion}

Not only does the Coulomb potential have very special relations with
$\sigma_{ltE^{-1}}^{2}$ scaling and phase space, it is also the only
mechanical potential which has been extended to a fully relativistic theory.
\ Thus the fundamental electromagnetic system of relativistic classical
electrodynamics is an elementary point charge $e$ of mass $m$ in a Coulomb
potential of strength $-e.$ \ In this case, the motion of the mechanical
system can be described by action angle variables $J_{1},$ $J_{2},$ $J_{3},$
with an energy\cite{Goldstein}
\begin{equation}
H=mc^{2}\left(  1+\frac{(e^{2}/c)^{2}}{\{J_{3}-J_{2}+[J_{2}^{2}-(e^{2}%
/c)^{2}]^{1/2}\}^{2}}\right)  ^{-1/2}%
\end{equation}
Under the single-parameter scaling $\sigma_{ltE^{-1}}$ which leaves classical
electromagnetism invariant, only the mass $m$ and energy $H$ are rescaled,
$m\rightarrow m/\sigma_{ltE^{-1}},$ $H\rightarrow H/\sigma_{ltE^{-1}}.$ \ The
action variables $J_{i}$ ( with units of $energy\times time$) as well as the
constants $e$ and $c$ are\ all invariant under the $\sigma_{ltE^{-1}}$ scaling
of lengths, times, and energies given in Eq.(1). \ Thus from Eq. (20), the
hamiltonian divided by the mass-energy, $H/mc^{2},$ is a dimensionless
function which describes the shape of the particle orbit as well as the speed
of the particle and is completely independent of any rescaling of the form
given by a dilatation factor $\sigma_{ltE^{-1}}$. \ The Coulomb potential
allows the decoupling of the phase-space behavior given by the $J_{i}$ from
the particle mass $m,$ and hence allows the possibility of a universal
spectrum of blackbody radiation in classical physics.

In the case of a restriction to circular orbits of angular momentum $J$,
$J_{1}=J_{2}=J_{3}=J,$ the energy becomes\cite{unfamiliar}
\begin{equation}
H=mc^{2}\left[  1-\left(  \frac{e^{2}}{Jc}\right)  ^{2}\right]  ^{1/2}%
\end{equation}
the orbital radius is%
\begin{equation}
r=\left(  \frac{e^{2}}{mc^{2}}\right)  \left(  \frac{Jc}{e^{2}}\right)
^{2}\left[  1-\left(  \frac{e^{2}}{Jc}\right)  ^{2}\right]  ^{1/2}%
\end{equation}
and the orbital frequency is
\begin{equation}
\omega=\left(  \frac{mc^{3}}{e^{2}}\right)  \left(  \frac{e^{2}}{Jc}\right)
^{3}\left[  1-\left(  \frac{e^{2}}{Jc}\right)  ^{2}\right]  ^{-1/2}%
\end{equation}
while the orbital speed is simply%
\begin{equation}
v=r\omega=e^{2}/J
\end{equation}
We note that Eq. (20) contains a singularity at $J_{2}=e^{2}/c.$ \ In the
circular orbit equations (21)-(24), the lower limit for the action variable
$J$ corresponds to $J\rightarrow e^{2}/c$ giving particle speed approaching
the speed of light $v\rightarrow c,$ frequency diverging $\omega
\rightarrow\infty,$ orbital radius approaching zero $r\rightarrow0,$ and total
energy going to zero $H\rightarrow0.$ \ We also emphasize that here for the
Coulomb potential (in contrast to all other mechanical potentials), the need
for relativity has nothing to do with the magnitude of the mass $m,$ \ but
rather is completely controlled by the ratio $J_{2}/(e^{2}/c).$ \ Furthermore,
for a relativistic particle $m$ in the Coulomb potential, the total energy $H$
in Eq. (20) for a bound particle takes on values between $0$ and $mc^{2}$;
thus a relativistic particle in a Coulomb potential can radiate away at most a
finite amount of energy $mc^{2}.$ \ In contrast, a nonrelativistic particle in
the Coulomb potential (where for a nonrelativistic circular orbit
$\mathcal{E}=H-mc^{2}\rightarrow-(1/2)m(e^{2}/J)^{2}$, $r\rightarrow
J^{2}/(me^{2}),$ $\omega\rightarrow me^{4}/J^{3},$ $v=e^{2}/J)$ can radiate
away an infinite amount of energy as $J\rightarrow0.$ \ In the nonrelativistic
limit, $J_{2}>>e^{2}/c\rightarrow0,$ the constant $e^{2}/c$ vanishes so that
now both $v$ and $J$ can assume all positive real values between $0$ and
$\infty.$ The nonrelativistic Coulomb problem is like the rest of
nonrelativistic mechanics in having no fundamental connection between energy
and time. \ 

\section{Matter-Radiation Connection}

\subsection{Equilibrium at Absolute Zero}

First we consider thermal equilibrium at the absolute zero of temperature
$T=0.$ \ We have seen above that the description of van der Waals forces
between macroscopic objects within classical physics requires the presence of
classical electromagnetic radiation with the Lorentz-invariant spectrum given
by Eq. (4) or (15). \ It seems natural to expect that a Lorentz-invariant
theory of classical electrodynamics would be required to accommodate a
Lorentz-invariant spectrum of random radiation, and we anticipate that an
elementary charge $e$ in a Coulomb potential $-e$ treated within relativistic
classical electron theory will leave invariant the Lorentz-invariant spectrum
of classical zero-point radiation. \ 

Indeed, the Coulomb potential has all the qualitatively correct aspects for
this invariance. \ If we consider an elementary charge $e$ of mass $m$ in a
circular Coulomb orbit described by angular momentum $J$, then from Eqs.
(22)-(24) the charge radiates a power $P_{rad}$ given by\cite{Jackson}
\begin{equation}
P_{rad}=\frac{2}{3}\frac{e^{2}}{c^{3}}\omega^{4}\gamma^{4}r^{2}=\frac{2}%
{3}(mc^{2})\left(  \frac{mc^{3}}{e^{2}}\right)  \left(  \frac{e^{2}}%
{cJ}\right)  ^{8}\left[  1-\left(  \frac{e^{2}}{cJ}\right)  ^{2}\right]  ^{-3}%
\end{equation}
or dividing by the characteristic energy per unit characteristic time%
\begin{equation}
\frac{P_{rad}}{(mc^{2})(mc^{3}/e^{2})}=\frac{2}{3}\left(  \frac{e^{2}}%
{cJ}\right)  ^{8}\left[  1-\left(  \frac{e^{2}}{cJ}\right)  ^{2}\right]  ^{-3}%
\end{equation}
The last equation is invariant under $\sigma_{ltE^{-1}}$ scaling and the
right-hand side depends only on the action variable $J.$ \ Furthermore the
particle will emit radiation into the harmonics of the frequency $\omega$ of
the mechanical motion. \ The radiation per unit solid angle emitted into the
$n$th harmonic is given by\cite{Jackson2}
\begin{equation}
\frac{dP_{rad}}{d\Omega}=\frac{e^{2}\omega^{4}r^{2}}{2\pi c^{3}}n^{2}\left\{
\left[  \frac{dJ_{n}(n\beta\sin\theta)}{d(n\beta\sin\theta)}\right]
^{2}+\frac{(\cot\theta)^{2}}{\beta^{2}}\left[  J_{n}(n\beta\sin\theta)\right]
^{2}\right\}
\end{equation}
The energy radiated into the $n$th harmonic has a multiplicative factor of
$n^{2}\omega^{4}r^{2}$ times a function of $n\beta.$ \ Since in the Coulomb
potential the circular orbital velocity is a function of the action variable
$J$ alone and is independent of the particle mass $m$, the \textit{ratios} of
energy radiated into the different harmonics depend only upon the action
variable $J$ (or equivalently on the particle velocity) and not on the mass
$m$. \ Indeed, the radiation into each of the individual modes associated with
the vector spherical multipole radiation has been calculated and has the same
behavior.\cite{Burko} \ 

Now in thermodynamic equilibrium, we expect the probability distribution for
action variables $J_{matter}$ of the matter to be related to the action
variables $J_{\lambdabar}$ of the radiation. For the harmonic dipole
oscillator of frequency $\omega_{0}$ evaluated in the infinite-mass
limit,\cite{harmonic} all the radiation is exchanged with the radiation modes
at the fundamental frequency $\omega_{0},$ and the phase space probability
distribution $P_{\omega_{0}}(J_{\omega_{0}})$ is exactly the same as that
given in Eq. (5) for the radiation mode of frequency $\omega_{0}%
$\cite{adiabatic}%
\begin{equation}
P_{\omega_{0}}(J_{\omega_{0}})=\frac{c}{b_{s}}\exp\left[  -\frac{J_{\omega
_{0}}c}{b_{s}}\right]
\end{equation}
Thus for the point harmonic oscillator, the phase space distribution $P(J)$
for matter is universal in the sense that it is independent of the
scale-giving parameter $\omega_{0}.$ \ Indeed, one can show that this holds
generally for point mechanical systems without harmonics.\cite{adiabatic}
\ However, all such systems are idealized in their interaction with radiation
because they have no finite size or velocity. \ 

In contrast with these idealized point systems with no harmonics, a charged
particle in the Coulomb potential represents a realistic mechanical system of
finite size and particle speed which allows this same separation of the phase
space from the scale-giving parameter $m$. \ For a particle of mass $m$ and
charge $e$ in the Coulomb potential, the probability distribution $P_{C}%
(J_{i},0)$ depends upon the exchange of energy with all the radiation modes at
frequencies $\omega_{n}=n\omega$ which are multiples of the fundamental
frequency $\omega;$ however, the \textit{ratios} of the power absorbed and
radiated in the $n$th harmonic compared to the fundamental $\omega$ are
completely independent of the mass $m$ or of the frequency of the fundamental.
\ But then the probability distribution $P_{C}(J_{i},0)$ for the action
variables $J_{i}$ of the matter will reflect information about the radiation
action variables $J_{\lambdabar}$ and will be independent of the particle mass
$m$. \ In addition to suggesting that the Lorentz-invariant and
scale-invariant zero-point radiation spectrum is invariant under scattering by
a charge $e$ in a Coulomb potential, this dependence of the $J_{i}$ on the
$J_{\lambdabar}$ alone fits exactly with our idea that at temperature $T=0$,
the particle should have a probability distribution which reflects zero
entropy and is independent of mass $m.$ \ Thus we have%
\begin{equation}
P_{C}(J_{i},0)=F\left(  \frac{J_{i}}{e^{2}/c},\frac{e^{2}}{b_{s}}\right)
\end{equation}
where $F$ is at present an unknown function. \ The entropy $S/k_{B}$ (divided
by Boltzmann's constant $k_{B})$ which follows from this probability
distribution on phase space would also be independent of any $\sigma
_{ltE^{-1}}$ scaling and independent of the mass $m$. \ 

These ideas hold not only for circular orbits but for arbitrary orbits for an
elementary charge in the Coulomb potential. \ In all cases, the radiation
balance at zero temperature and the probability distribution for the action
variables is independent of the particle mass $m$. \ Indeed, classical
electromagnetic zero-point radiation is both scale invariant and conformal
invariant. \ The only available scale is given by the particle mass $m$. \ For
a relativistic Coulomb potential, considerations of scaling alone dictate that
the mass $m$ can not enter the probability distribution at zero temperature.
\ \ This is a first crucial step in understanding how the thermal radiation
pattern can be universal within classical physics despite the interaction with
matter involving various masses. \ It is the crucial separation of the mass
parameter from the underlying conformal structure which should make possible a
universal radiation equilibrium within relativistic classical
physics.\cite{fine}

\subsection{Equilibrium at Finite Temperature}

If a charged particle in the Coulomb potential is placed in a large container
with conducting walls where there is a finite amount of energy above the
zero-point radiation, then the relativistic Coulomb system will scatter the
radiation and presumably produce a state of thermal equilibrium. \ In thermal
equilibrium, the available energy $U$ above the zero-point energy has been
shared between the scattering system and the thermal radiation in the
container. \ If the container is large enough, then the scattering system will
absorb a negligible fraction of the available energy $U.$ \ In equilibrium, we
do not expect that the scattering system will respond to the\ container's
total energy $U$ (the extensive variable), but rather to the local energy per
unit volume $U/V$ (the intensive variable). \ Under dilatation, this energy
per unit volume scales with four powers of the scaling parameter
$\sigma_{ltE^{-1}},$ one power of $\sigma_{ltE^{-1}}$ coming from the energy
and three more powers of $\sigma_{ltE^{-1}}$ from the volume. \ Thus if we
want to obtain an energy which scales with one power of $\sigma_{ltE^{-1}}$
and is an intensive variable, then we must choose $[U/(Va_{S}/k_{B}%
^{4})]^{1/4}$ where $a_{s}$ is Stefan's constant$.$ \ Of course, this
corresponds exactly to $k_{B}T=[U/(Va_{S}/k_{B}^{4})]^{1/4}$ from Stefan's
law. Now the probability distribution $P_{\lambdabar}(J_{\lambdabar
},\lambdabar k_{B}T/b_{s})$ for the action variable $J_{\lambdabar}$ of the
radiation mode of wavelength $\lambdabar$ is given by Eq. (7). \ The
probability distribution is no longer the same for all wavelengths
$\lambdabar,$ but rather involves a function of the dimensionless quantity
$\lambdabar k_{B}T/b_{s}$ connecting the wavelength of the radiation mode and
the energy $k_{B}T$ associated with the thermal radiation. \ Since the
frequency of the radiation mode is directly connected to the wavelength,
$\omega=c/\lambdabar,$ the functional dependence could just as well be
expressed as involving ($b_{s}/c)\omega/(k_{B}T).$\ \ 

The scattering situation at finite temperature is very similar to that at
zero-temperature. \ For an elementary charge $e$ in a Coulomb potential, the
only parameter which scales with the the dilatation factor $\sigma
_{l,t,E^{-1}}$ of electromagnetism is the particle mass $m.$ \ Thus comparing
matter and thermal radiation, the ratio of characteristic energies involves
$mc^{2}/(k_{B}T),$ the ratio of characteristic lengths involves [$e^{2}%
/(mc^{2})]/[b_{s}/k_{B}T],$ and the ratio of characteristic times involves
[$e^{2}/(mc^{3})]/[b_{s}/(ck_{B}T)].$ \ Every one of these ratios involves
universal constants multiplying the dimensionless ratio $k_{B}T/(mc^{2}%
)=[U/(Va_{S}/k_{B})]^{1/4}/(mc^{2})$ or its inverse$.$ Thus the probability
distribution for the action variables $J_{i}$ of the mechanical system of mass
$m$ is of the form%

\begin{equation}
P_{C}(J_{i},k_{B}T/mc^{2})=G\left(  \frac{J_{i}}{e^{2}/c},\frac{e^{2}}{b_{s}%
},\frac{k_{B}T}{mc^{2}}\right)
\end{equation}
where $G$ is at present an unknown function which goes over to the unknown
function $F\left(  J_{i}/(e^{2}/c),e^{2}/b_{s}\right)  $ in Eq. (29) when
$T=0$. \ The crucial thing here is that the mass $m$ enters Eq. (30) only in
the ratio $k_{B}T/(mc^{2}).$

We should note the crucial parallel between radiation and matter which exists
in relativistic classical electron theory with a discrete charge. \ Thermal
radiation at absolute zero involves the same distribution\ (5) in the action
variables $J_{\lambdabar}$ for each radiation mode $\lambdabar,$ independent
of the mode wavelength. \ In parallel fashion, hydrogen-like Coulomb
scattering systems at absolute zero also all involve the same distribution
(29) in their action variables $J_{i}$ independent of the mass $m.$ \ However,
the average energy $U_{\lambdabar}$ for each radiation mode is different,
$U_{\lambdabar}=<J_{\lambdabar}>c/\lambdabar=<J_{\lambdabar}>\omega
_{\lambdabar},$ because the energy $U_{\lambdabar}$ involves both the
scale-giving parameter $\lambdabar$ and the average over the distribution of
action variables $J_{\lambdabar},$ while the energy $U_{m}$ for each Coulomb
system is different, $U_{m}=mc^{2}<f(J_{i})>,$ because the energy $U_{m}$
involves both the scale-giving parameter $m$ times an average over the
distribution of action variables. \ In thermal radiation, the $J_{\lambdabar}%
$-distribution for radiation depends on $(b_{s}/c)\omega_{\lambdabar}%
/k_{B}T=b_{s}/(\lambdabar k_{B}T)$ and the $J_{i}$-distribution for matter
depends upon $mc^{2}/(k_{B}T);$ in each case the single scale-giving parameter
($\omega=c/\lambdabar$ or $m)$ of the system is related to the temperature
$T.$

We should emphasize that, in a Coulomb potential, large and small masses act
differently in thermal radiation because they couple to different radiation
modes. \ Thus a small mass $m$ is coupled to low-frequency radiation modes
since for fixed action variables $J_{i},$ the frequencies $\omega_{i}$ (for
example, in Eq. (23)) vary directly as the particle mass $m.$ \ It is these
low frequencies where the thermal radiation predominates. \ On the other hand,
a large mass $m$ is associated with high frequencies where the zero-point
radiation dominates, and the thermal radiation will have little influence on
the particle's motion in its Coulomb orbit. \ Thus for particles in a Coulomb
potential in thermal radiation, there is a transition which can be associated
with the particle mass $m$, just as as there is a transition in the radiation
modes which can be associated with the frequency $\omega_{\lambdabar}$ of the
radiation modes. \ We notice that this situation is completely different from
that of a nonrelativistic mass in a harmonic potential well where the natural
oscillation frequency is given by $\omega_{0}=(K/m)^{1/2},$ and, for fixed
spring constant $K,$ \textit{decreases} with increasing mass. \ 

The situation for radiation equilibrium is enormously simplified for the
relativistic Coulomb potential. \ Thus if radiation equilibrium exists for one
mass $m$ in a Coulomb potential in zero-point radiation, then it exists for
all masses $m$ since the phase space distribution $P_{C}(J_{i},0)$ must be
independent of $m,$ and the interaction with radiation is determined by the
phase space distribution. \ Furthermore, if equilibrium exists for one mass
$m$ in the Planck spectrum (6) at all temperatures $T,$ then the equilibrium
is valid for all masses $m$ since the phase space distribution $P_{C}%
(J_{i},k_{B}T/mc^{2})$ depends only upon the ratio $k_{B}T/(mc^{2}).$. \ This
simplification will not hold for any non-Coulomb potential function. \ 

\section{Discussion}

Despite all the claims to the contrary, the blackbody radiation problem is
still an unsolved problem within classical physics. \ Today, the textbooks and
orthodox physics literature claim that classical physics can not explain the
observed Planck spectrum of blackbody radiation. \ Indeed it was the apparent
inability of classical physics to account for this spectrum which led to the
introduction of quantum theory in the years after 1900. \ We believe that the
perspective on blackbody radiation current in the physics community today
arises because the physicists at the beginning of the twentieth century missed
three essential aspects of classical electrodynamics. \ 

First, they were unaware of classical electromagnetic zero-point radiation.
Classical electromagnetic zero-point radiation enters classical
electromagnetic theory as the homogeneous boundary condition on Maxwell's
equations and is required to account for the experimentally observed van der
Waals forces between macroscopic objects.\cite{Sparnaay} \ However, in the
research of the early twentieth century, there were no direct measurements of
van der Waals forces and the homogeneous solution of Maxwell's equations was
taken to vanish. \ In his work on classical electron theory, Lorentz
specifically assumes\cite{boundary} that all radiation arises at finite time;
there is no classical zero-point radiation. \ Historically, zero-point
radiation entered physics only after the advent of quantum theory and even
today plays an ambiguous role. \ Some physicists are sure that any idea of
zero-point energy must involve quantum mechanics. \ It is only in the second
half of the twentieth century that \textit{classical} electromagnetic
zero-point radiation was developed in the classical description of
nature.\cite{delaP}\cite{dcC}

Second, the physicists at the beginning of the twentieth century did not take
seriously the requirements of special relativity. \ \ Special relativity was
(and still is) regarded as a specialty subject which needs to be considered
only for high-speed particles. \ Thus Lorentz's classical electron theory
involved point charges in nonrelativistic potentials and discussions of atomic
physics were all in the context of nonrelativistic mechanics. \ Indeed quantum
mechanics was developed as a subterfuge to fix the connection between
nonrelativistic mechanics and electromagnetic radiation, and
Heisenberg-Schroedinger quantum mechanics remains a nonrelativistic theory to
the present day. \ In this same vein, even the physicists who recently
developed the ideas of classical electron theory with classical
electromagnetic zero-point radiation\cite{delaP} into a theory designated as
stochastic electrodynamics, SED, failed to appreciate the importance of
relativity for the mechanical scattering systems. \ Thus, for example, a
review article in 1975 considers arbitrary nonrelativistic potentials and
states that Newton's second law in the nonrelativistic form $\mathbf{F}%
=m\mathbf{a}$ is to be used in the analysis of particles interacting with
classical electromagnetic zero-point radiation.\cite{Boyer1975}

Third, physicists have continued the nineteenth century conception of a
continuous scale of electric charge, repeatedly dealing with small electric
charges in connecting radiation and matter, rather than exploring the
implications of the discrete charges in nature. \ Thus Planck considered
harmonic oscillators in the walls of the blackbody cavity, but the charge on
the oscillators was merely regarded as very small and cancelled out in his
final result. \ It is true that in the early years of the twentieth century
Planck did hope to connect the electron charge $e$ to his own constant $h$
since he noted that $e^{2}/c$ and $h$ have the same units.\cite{Kuhn2}
\ However, Lorentz, even in his account of classical electron theory in 1915,
is still looking for a constant analogous to Planck's constant without
remarking that $e^{2}/c$ provides just such a constant.\cite{Lorentz}
\ Lorentz does not note the suitability of $e^{2}/c$ because he is looking
within nonrelativistic classical mechanics, not within classical
electrodynamics. \ Furthermore, in the last third of the twentieth century,
the developers of classical electron theory with classical electromagnetic
zero-point radiation always have taken the small charge limit so as to deal
with quasi-Markov stochastic processes for matter.\cite{delaP}

The blackbody radiation problem within classical physics has been explored
repeatedly. \ In addition to the old derivations of the Rayleigh-Jeans
spectrum given by the renowned physicists at the beginning of the twentieth
century, there have also been repeated derivations of Planck's spectrum. \ In
the presence of classical zero-point radiation, the Planck spectrum within
classical physics has been derived using various ideas of classical physics:
energy equipartition of nonrelativistic translational degrees of freedom in
the large mass limit,\cite{equi} thermal fluctuations above zero-point
radiation,\cite{fluct} comparisons between diamagnetic and paramagnetic
behavior,\cite{interpol} the acceleration of point electromagnetic systems
through zero-point radiation,\cite{thermal2}\cite{thermal3}\cite{thermal4} and
entropy ideas connected with Casimir forces.\cite{conjectured} \ \ Most of
these derivations involve harmonic oscillator-like systems which interact with
radiation at a single frequency in the infinite-mass-and-zero-velocity limit.
\ All show a natural connection between classical electromagnetic zero-point
radiation and Planck's spectrum of thermal radiation. \ None involves a full
relativistic scattering calculation. \ However, all the insights of these
derivations have been rejected by physicists who insist on the validity of the
scattering calculations which have used nonrelativistic, nonlinear, mechanical
systems\cite{nonlinear} to scatter classical electromagnetic zero-point
radiation toward the Rayleigh-Jeans spectrum. \ Despite the fact that one
might expect only a relativistic scattering system to maintain the invariance
of the Lorentz-invariant spectrum of classical electromagnetic radiation, most
physicists are so confident of the universal applicability of nonrelativistic
physics that they find it hard to conceive of the possibility that relativity
might be required for appropriate scatterers for thermal
radiation.\cite{Blanco} \ However, the thermal effects of acceleration through
the Lorentz-invariant spectrum of classical zero-point radiation point
unambiguously in the direction of a relativistic theory for blackbody
radiation. And a relativistic scattering calculation using the Coulomb
potential has never been done but has all the qualitative aspects appropriate
to blackbody equilibrium. \ 

It is our guess that (just as is found from uniform acceleration through the
classical zero-point vacuum) the blackbody radiation spectrum has nothing to
do with energy quanta and everything to do with the conformal symmetries of
classical electromagnetism.\ In the present work, we have tried to suggest why
relativity and discrete electric charge are probably crucial to understanding
blackbody radiation within classical electromagnetic theory. \ Indeed,
zero-point radiation, relativity, and discrete charge within classical physics
are probably crucial to a deeper understanding of much more of atomic and
statistical physics.


\begin{thebibliography}{99}                                                                                               %


\bibitem {thermal}Thermal effects of acceleration appeared first in work on
quantum field theory by P. C. Davies, "Scalar particle production in
Schwarzschild and Rindler metrics," J. Phys. A \textbf{8}, 609-616 (1975) and
by W. G. Unruh, "Notes on black hole evaporation," Phys. Rev. D \textbf{14},
870-892 (1976). \ 

\bibitem {thermal2}T. H. Boyer, "Thermal effects of acceleration through
random classical radiation," Phys. Rev. D \textbf{21}, 2137-2148 (1980);
"Thermal effects of acceleration for a classical dipole oscillator in
classical electromagnetic zero-point radiation," Phys. Rev. D \textbf{29},
1089-1095 (1984).

\bibitem {thermal3}T. H. Boyer, "Derivation of the blackbody radiation
spectrum from the equivalence principle in classical physics with classical
electromagnetic zero-point radiation," Phys. Rev. D \textbf{29}, 1096-1098
(1984). \ "Thermal effects of acceleration for a classical spinning magnetic
dipole in classical electromagnetic zero-point radiation," Phys. Rev. D
\textbf{30}, 1228-1232 (1984). \ 

\bibitem {thermal4}D. C. Cole, "Properties of a classical charged harmonic
oscillator accelerated through classical electromagnetic zero-point
radiation," Phys. Rev. D \textbf{31}, 1972-1981 (1985).

\bibitem {conform}Some aspects of the symmetries have been given by J.
Haantjes, "Die Gleichberechtigung gleichformig beschleunigter Beobacter
f\"{u}r die electromagnetischen Erscheingungen," Koninkl. Ned. Akad.
Wetenschap. Proc. \textbf{43}, 1288-1299 (1940) and by\ T. H. Boyer,
"Blackbody radiation, conformal symmetry, and the mismatch between classical
mechanics and electromagnetism," J. Phys. A \textbf{38}, 1807-18221 (2005).

\bibitem {Eisberg}See, for example, R. Eisberg and R. Resnick, \textit{Quantum
Physics of Atoms, Molecules, Solids, Nuclei, and Particles}, 2nd ed, (Wiley,
New York 1985), p. 12.

\bibitem {Lavenda}See, for example, B. H. Lavenda, \textit{Statistical
Physics: A Probabilistic Approach}, (Wiley, New York 1991), pp. 73-74, and F.
Reif, \textit{Fundamentals of Statistical and Thermal Physics}, (McGraw-Hill,
New York 1965), pp. 251-252.

\bibitem {nonlinear}J. H. Van Vleck, "The Absorption of Radiation by Multiply
Periodic Orbits, and its Relation to the Correspondence Principle and the
Rayleigh-Jeans Law. Part II Calculation of Absorption by Multiply Periodic
Orbits," Phys. Rev. \textbf{24}, 347-365 (1924). \ T. H. Boyer, "Equilibrium
of random classical electromagnetic radiation in the presence of a
nonrelativistic nonlinear electric dipole oscillator," Phys. Rev. D
\textbf{13}, 2832-2845 (1976). \ T. H. Boyer, "Statistical equilibrium of
nonrelativistic multiply periodic classical systems and random classical
electromagnetic radiation," Phys. Rev. A \textbf{18}, 1228-1237 (1978).

\bibitem {Kuhn}T. S. Kuhn, \textit{Black-Body Theory and the Quantum
Discontinuity 1894-1912}, (Oxford U. Press, New York 1978), Chapter VIII.

\bibitem {Marshall}T. W. Marshall, "Statistical Electrodynamics," Proc. Camb.
Phil. Soc. \textbf{61}, 537-546 (1965); T. H. Boyer, "Derivation of the
Blackbody Radiation Spectrum without Quantum Assumptions," Phys. Rev.
\textbf{182}, 1374-1383 (1969).

\bibitem {Boyer1975}T. H. Boyer, "Random electrodynamics: The theory of
classical electrodynamics with classical electromagnetic zero-point
radiation," Phys. Rev. D \textbf{11}, 790-808 (1975).

\bibitem {Sparnaay}M. J. Sparnaay, "Measurement of the attractive forces
between flat plates," Physica \textbf{24}, 751-764 (1958). \ S. K. Lamoreaux,
"Demonstration of the Casimir force in the 0.6 to 6 $\mu$m range," Phys. Rev.
Lett. \textbf{78}, 5-8 (1997); \textbf{81}, 5475-5476 (1998). \ U. Mohideen,
"Precision measurement of the Casimir force from 0.1 to 0.9 $\mu$m," Phys.
Rev. Lett. \textbf{81}, 4549-4552 (1998). \ H. B. Chan, V. A. Aksyuk, R. N.
Kleiman, and F. Capasso, "Quantum mechanical actuation of
microelectromechanical systems by the Casimir force," Science \textbf{291},
1941-1944 (2001). \ G. Bressi, G. Carugno, R. Onofrio, and G. Ruoso,
"Measurement of the Casimir force between parallel metallic surfaces," Phys.
Rev. Lett. \textbf{88}, 041804(4) (2002). 

\bibitem {Casimirmodel}The analogy between the electrostatic scaling behavior
and the electromagnetic scaling behavior provides the basis for Casimir's
model of the electron;\ H. B. G. Casimir, "Introductory remarks on quantum
electrodynamics," Physica \textbf{19}, 846-849 (1956). \ It was not realized
that the scaling analogy holds very generally.

\bibitem {Casimir}The parallel-plate calculation was first made by H. B. G.
Casimir, "On the attraction between two perfectly conducting plates," Koninkl.
Ned. Akad. Wetenschap, Proc. \textbf{51}, 793-795 (1948). \ 

\bibitem {sphere}\ The spherical case was first calculated by T. H. Boyer,
"Quantum electromagnetic zero-point energy of a conducting spherical shell and
the Casimir model for a charged particle," Phys. Rev. \textbf{174}, 1764-1776 (1968).

\bibitem {hbarc}Because of historical accident, Stefan's constant is usually
rewritten using $a_{s}=k_{B}^{4}\pi^{2}/(15\hbar^{3}c^{3}).$ \ See, for
example, P. H. Morse, \textit{Thermal Physics, 2nd ed.}, (Benjamin/Cummings
Publishing, Reading, Mass. 1969), p. 339. \ Then the dimensionless universal
constant is given as $e^{2}(a_{S}/k_{B}^{4})^{1/3}=e^{2}\{[\pi^{2}k_{B}%
^{4}/(15\hbar^{3}c^{3})]/k_{B}^{4}\}^{1/3}=(\pi^{2}/15)[e^{2}/(\hbar
c)]=0.00063487$. \ Thus we recognize the dimensionless constant as related to
Sommerfeld's fine-structure constant. \ However, for many physicists
(including both the physicists of the early twentieth century and also recent
referees), the use of Planck's constant $\hbar$ rather than Stefan's constant
$a_{s}$ serves as an overwhelming distraction. The appearance of Planck's
constant seems to cause an immediate fixation on ideas of energy quanta.
\ There are no ideas of energy quanta in the present analysis, and such ideas
have no place in classical physics. \ Actually, Planck's constant does not
necessarily have any connection to energy quanta; it can serve as the scale
factor for classical electromagnetic zero point radiation without in any way
implying energy quanta. See, for example, ref. 11.

\bibitem {bs}Readers will recognize $b_{s}$ in terms of currently familiar
constants as $b_{s}=(1/2)\hbar c.$ \ We emphasize that no energy quanta
whatsoever are used in our classical analysis.

\bibitem {temperature}See, for example, L. L. Henry and T. W. Marshall, "A
classical treatment of dispersion forces," Nuovo Cimento \textbf{41}, 188-197
(1966); T. H. Boyer, Van der Waals forces and zero-point energy for dielectric
and permeable materials," Phys. Rev. A \textbf{9}, 2078-2084 (1974).

\bibitem {Rice}S. O. Rice, in \textit{Selected Papers on Noise and Stochastic
Processes}, edited\ by N. Wax (Dover, New York, 1954), p. 133.

\bibitem {Power}See, for example, E. A. Power, \textit{Introductory Quantum
Electrodynamics }(American Elsevier, New York 1964), pp. 18-22.

\bibitem {adiabatic}T. H. Boyer, "Connection between the adiabatic hypothesis
of old quantum theory and classical electrodynamics with classical
electromagnetic zero-point radiation," Phys. Rev. A \textbf{18}, 1238-1245
(1975). \ 

\bibitem {entropy}The existence of energy fluctuations which involve no
entropy requires ideas of special relativity and is foreign to nonrelativistic
classical statistical mechanics.

\bibitem {gamma}E. Cunningham, "The Principle of Relativity in Electrodynamics
and an Extension Thereof," Proc. London Math. Soc. \textbf{8}, 77-98 (1910);
H. Bateman, "The Transformation of the Electrodynamical Equations," Proc.
London Math. Soc. \textbf{8}, 223-264 (1910).

\bibitem {conformal}T. H. Boyer, "Conformal symmetry of classical
electromagnetic zero-point radiation," Found. Phys. \textbf{19}, 349-365
(1989), Eq. (40).

\bibitem {RJ}The correlation function can be obtained by following the same
pattern as given in ref. 24 except that we replace $(1/2)\hbar c\left\vert
k\right\vert $ by $k_{B}T$ in Eq. (39). \ The integral then involves $%
%TCIMACRO{\dint }%
%BeginExpansion
{\displaystyle\int}
%EndExpansion
(d^{3}k/\left\vert k\right\vert \cos[\mathbf{k}\cdot(\mathbf{r-r}^{\prime
})-c\left\vert k\right\vert (t-t^{\prime})]=%
%TCIMACRO{\dint \limits_{0}^{\infty}}%
%BeginExpansion
{\displaystyle\int\limits_{0}^{\infty}}
%EndExpansion
dk%
%TCIMACRO{\dint \limits_{0}^{\pi}}%
%BeginExpansion
{\displaystyle\int\limits_{0}^{\pi}}
%EndExpansion
d\theta\sin\theta%
%TCIMACRO{\dint \limits_{0}^{2\pi}}%
%BeginExpansion
{\displaystyle\int\limits_{0}^{2\pi}}
%EndExpansion
d\phi\cos[k(R\cos\theta-c\tau)]=%
%TCIMACRO{\dint \limits_{0}^{\infty}}%
%BeginExpansion
{\displaystyle\int\limits_{0}^{\infty}}
%EndExpansion
(dk/k)\{\sin[k(R-c\tau)]+\sin[k(R+c\tau)]\}.$ \ Finally the definite integral
gives $%
%TCIMACRO{\dint \limits_{0}^{\infty}}%
%BeginExpansion
{\displaystyle\int\limits_{0}^{\infty}}
%EndExpansion
dx(\sin mx)/x=\pi/2$ if $m>0$, $0$ if $m=0$, and $-\pi/2$ if $m<0.$

\bibitem {conjectured}T. H. Boyer, "Conjectured derivation of the Planck
radiation spectrum from Casimir energies," J. Phys. A: Math. Gen. \textbf{36},
7425-7440 (2003), Section 6.

\bibitem {quandry}As far as the quantum theorists are concerned, the
introduction of quanta solved the blackbody problem in conjunction with
statistical mechanics. \ Thus they are surprised when the Planck spectrum
appears without the use of quantum statistics. \ The point of view presented
here is entirely classical and is quite different. \ In the classical
perspective present here, classical statistical mechanics plays no role but
rather is a derived concept, derived from the randomness of random phases of
waves, in the sense of ref. 19. \ The connection between conformal motions and
gravity is crucial in connecting the Planck spectrum to thermal radiation.
\ Thus if thermal radiation exists in equilibrium in a gravitational field, it
must correspond to the radiation associated with uniform acceleration through
zero-point radiation plus perhaps additional radiation. \ The finite
temperature radiation must fit on top of the spectrum from acceleration
through zero-point radiation. \ In order to satisfy Wien's displacement
theorem, the additional radiation must follow the same pattern which already
appeared from the acceleration through classical zero-point radiation.

\bibitem {Rindler}W. Rindler, \textit{Essential Relativity: Special, General,
and Cosmological, }2nd ed. (Springer-Verlag, New York1977), pp 49-51.

\bibitem {harmonic}During the early years of the twentieth century, the
nonrelativistic harmonic oscillator was often connected to radiation by a very
small charge $q.$ \ See, for example, M. Planck, \textit{The Theory of Heat
Radiation} (Dover, New York 1959). \ The electric dipole oscillator was said
to come to equilibrium with an energy $U$ equal to the energy $U_{\omega_{0}}$
of the radiation mode at the same frequency as the natural frequency of the
oscillator, $U=U_{\omega_{0}}.$ \ This equality held for fixed $\omega_{0}$
whether the mass $m$ and spring constant $K$ of the oscillator were both large
and the velocity and amplitude were both small, or whether $m$ and $K$ were
both small with the velocity and amplitude large, perhaps the velocity even
exceeding $c$. \ However, classical electromagnetic radiation does not treat
these two possible radiating systems alike because finite velocity and
amplitude means that radiation will be emitted into the higher harmonics of
the fundamental frequency. \ The traditional classical calculations are
accurate only in the limit $m\rightarrow\infty$ where the velocity and
amplitude vanish; in this limit, the oscillator interacts with radiation only
at its fundamental frequency and so does not scatter radiation toward equilibrium.

\bibitem {ponderable} H. A. Lorentz, \textit{The Theory of Electrons and its
Application to the Phenomena of Light and Radiation Heat}, 2nd ed. (Dover, New
York 1952), p. 96. \ This is a republication of the 2nd edition of 1915.

\bibitem {scal}T. H. Boyer, "Scaling Symmetry and Thermodynamic Equilibrium
for Classical Electromagnetic Radiation," Found. Phys. \textbf{19}, 1371-1383
(1989). \ D. C. Cole, "Classical Electrodynamic Systems Interacting with
Classical Electromagnetic Random Radiation," Found. Phys. \textbf{20}, 225-240 (1989).

\bibitem {Goldstein}See, for example, H. Goldstein, \textit{Classical
Mechanics, 2nd ed.,} (Addison-Wesley Publishing, Reading, Mass. 1981), p. 498.
We are using action variables which are $1/(2\pi)$ times those in Goldstein's text.

\bibitem {unfamiliar}T. H. Boyer, "Unfamiliar trajectories for a relativistic
particle in a Kepler or Coulomb potential," Am. J. Phys. \textbf{72}, 992-997 (2004).

\bibitem {Jackson}J. D. Jackson, \textit{Classical Electrodynamics, 2nd. ed.},
(Wiley, New York 1975), p. 664.

\bibitem {Jackson2}See ref. 34, p. 695.

\bibitem {Burko}L. M. Burko, "Self-force approach to synchrotron radiation,"
Am. J. Phys. \textbf{68}, 456-468 (2000).

\bibitem {fine}A complete classical electromagnetic calculation of radiation
equilibrium for a relativistic particle in the Coulomb potential at zero
temperature should give the value of the fine structure constant. \ The
probability distribution for the action variables $J_{i}$ involves two
different scales; \ $e^{2}/c$ and $b_{s}/c.$ \ The scale $e^{2}/c$ is related
to the relativistic mechanical behavior and provides a limit at small $J_{2}.$
\ See ref. 33. \ The scale $b_{s}/c$ involves the energy balance with the
random zero-point radiation. \ Roughly, when the action variables $J_{i}$ are
larger than $b_{s}/c,$ then the hydrogen system is losing more energy by
radiation emission than it is picking up from the zero-point radiation; when
the $J_{i}$ are smaller than $b_{s}/c,$ the reverse is true. \ See section D2
of ref. 11. \ We note that if $b_{s}/c$ were too small, $b_{s}/c<e^{2}/c,$
then no hydrogen atom would exist because the energy-conserving orbits at this
angular momentum would spiral into the nucleus. \ See ref. 33. \ In nature,
the value is $b_{s}/c\approx68e^{2}/c.$

\bibitem {boundary}See ref. 30, note 6, p. 240, which gives Lorentz's explicit
assumption on the boundary condition.

\bibitem {delaP}A review of most of the work on classical zero-point radiation
is presented by L. de la Pena and A. M. Cetto, \textit{The Quantum Dice - An
Introduction to Stochastic Electrodynamics }(Kluwer Academic, Dordrecht 1996).

\bibitem {dcC}D. C. Cole and Y. Zou, "Quantum Mechanical Ground State of
Hydrogen Obtained from Classical Electrodynamics," Phys. Letters A
\textbf{317}, 14-20 (2003). \ Cole and Zou's numerical simulation calculations
suggest that classical zero-point radiation acting on the classical model for
hydrogen (a point charge in a Coulomb potential) produces a steady-state
probability distribution which is finite and which seems to approach the
ground state obtained from solution of the Schroedinger equation for a point
charge in the Coulomb potential. Cole and Zou have no adjustable parameters in
their calculations. \ See also, T. H. Boyer, "Comments on Cole and Zou's
calculation of the hydrogen ground state in classical physics," Found. Phys.
Letters \textbf{16}, 613-617 (2003). \ 

\bibitem {Kuhn2}See ref. (Kuhn) p. 132.

\bibitem {Lorentz}On page 78 of ref. 30, Lorentz writes, \ "Now, if the state
of radiation is produced by a ponderable body, the values of the two constants
[in the blackbody spectrum] must be determined by something in the
constitution of this body, and these values can only have the universal
meaning of which we have spoken, if all ponderable bodies have something in common."

\bibitem {equi}T. H. Boyer, "Derivation of the Blackbody Radiation Spectrum
without Quantum Assumptions," Phys. Rev. \textbf{182}, 1374-1383 (1969); T. W.
Marshall, "Brownian motion of a mirror," Phys. Rev. D \textbf{24}, 1509-1515 (1981).

\bibitem {fluct}T. H. Boyer, "Classical Statistical Thermodynamics and
Electromagnetic Zero-Point Radiation," Phys. Rev. \textbf{186}, 1304-1318 (1969).

\bibitem {interpol}T. H. Boyer, "Derivation of the Planck radiation spectrum
as an interpolation formula in classical electrodynamics with classical
electromagnetic zero-point radiation," Phys. Rev. D \textbf{27}, 2906-2911
(1983); "Reply to 'Comment on Boyer's derivation of the Planck spectrum,'"
Phys. Rev. D \textbf{29}, 2418-2419 (1984). \ 

\bibitem {Blanco}See R. Blanco, L. Pesquera, and E. Santos, "Equilibrium
between radiation and matter for classical relativistic multiperiodic systems.
Derivation of Maxwell-Boltzmann distribution from Rayleigh-Jeans spectrum,"
Phys. Rev. D \textbf{27}, 1254-1287 (1983); "Equilibrium between radiation and
matter for classical relativistic multiperiodic systems II. Study of radiative
equilibrium with Rayleigh-Jeans radiation," Phys. Rev. D \textbf{29},
2240-2254 (1984). \ These articles suggest that a relativistic classical
scatterer again leads to the Rayleigh-Jeans spectrum. \ The calculations
involve a general class of potentials. \ The mechanical momentum of the
particle is calculated relativistically but the class of potentials excludes
the Coulomb potential and all purely electromagnetic scattering systems.
\ Many physicists do not realize that such mechanical systems do not satisfy
Lorentz invariance. \ It is the Coulomb potential and only the Coulomb
potential which has been incorporated into a fully relativistic classical
electron theory. \ A general potential (unextended to involve new
acceleration-dependent forces and new radiation specific to the potential)
violates the center-of-energy conservation laws which are directly related to
the generator of Lorentz transformations. \ See T. H. Boyer, "Illustrations of
the relativistic conservation law for the center of energy," Am. J.
Phys.\textbf{ 73}, 953-961 (2005).
\end{thebibliography}
\end{document}